\documentstyle[aps,twocolumn,psfig]{revtex}
\tightenlines
\begin{document}
\draft
\title{EQUATION OF STATE OF HADRONIC MATTER AND ELECTROMAGNETIC RADIATION FROM 
 RELATIVISTIC HEAVY ION COLLISIONS}
\author{ Jean Cleymans} 
\address{ Department of Physics, University of Cape Town, Rondebosch 7700,
South Africa }
\author{Krzysztof Redlich\cite{add}}
\address{ Gesellschaft f\"ur Schwerionenforschung (GSI),
D-64220 Darmstadt, Germany}

\author{ Dinesh  Kumar Srivastava }
\address{Variable Energy Cyclotron Centre, 1/AF Bidhan Nagar, Calcutta
700 064, India}
\date{\today}
\parindent=20pt
\maketitle
\begin{abstract}
We study the radiation of thermal photons and dileptons likely to be produced
in relativistic heavy ion collisions.
We find that the thermal photon multiplicity scales with the charged 
pion multiplicity  as $(dN_{ch}/dy)^\alpha$ with 
$\alpha\sim  1.2$ for a transversely expanding system,
 contrary to the general belief of a quadratic dependence. 
 The scaling is shown to be valid, both for real and virtual photons.
The coefficient of proportionality at a  given energy may
help us identify the appropriate equation of state of hot hadronic matter
produced in such collisions.
\end{abstract}

\narrowtext

The search for the quark-gluon plasma (QGP), 
a deconfined state of strongly
interacting matter, is entering a decisive phase.
Evidence has been mounting for some time now, that such  
matter is perhaps being produced in  collisions
involving heavy nuclei at the CERN 
Super Proton Synchrotron (SPS). If confirmed, this holds out a promise
that the QGP likely to be created at 
the Brookhaven Relativistic Heavy Ion Collider (RHIC) and the CERN
Large Hadron Collider (LHC) will occupy a much
larger space-time volume, in similar collisions. The plasma thus produced will
have to expand and cool, due to large internal pressure.
It has also to undergo hadronization, once the temperature drops below the
transition temperature. These hadrons may continue to interact for some time
before their density gets too low to support frequent collisions.
We may also have a mixed phase of quarks, gluons, and hadrons during which
the speed of sound may become vanishingly small, if quantum chromodynamics
admits a first order phase transition.

Many of the observables from the above collisions, like the particle
distributions, electromagnetic radiation, the spatial and temporal 
dimensions of the fireball etc., will be decided by the equation of state
 of hadronic matter. The transverse expansion of the system which 
drastically alters the time scales in the system, at least with regard
to the mixed and the hadronic phases is also affected by the equation
of state \cite{klrt}.  
What should be the equation of state of hot hadronic matter which is created
in heavy ion collisions?
It is expected that if the transition temperature is very close to 
the pion mass
then the hadronic matter could  possibly well be determined by pions and
only a few low lying mesonic resonances.
 The hadronic matter is likely to be populated by heavier hadrons once the
 temperature is higher.  Will these hadrons stay in chemical and thermal
equilibrium, till the freeze-out? This will be decided by the competition 
between the collision frequency and the rate of expansion, and if
valid, it will be reflected in uniquely determined particle ratios \cite{cs}.
It is not difficult to imagine that the equation of state or at least the
composition of the hadronic matter may  be  somehow related to the
manner in which the hadronization proceeds.
 This aspect has not yet received enough attention
as it involves the non-perturbative realm  of quantum chromodynamics.

The question of hadronization of the plasma gets very complex in view of the
recent findings that the QGP formed at RHIC or at LHC will not reach
chemical equilibrium during the life-time of the quark phase \cite{chem}.
We can understand this as follows. Recall that according to the
Maxwell criterion, the phase transition temperature ($T_c$) is given by
$P_Q(T_c)=P_H(T_c)$, where $P$ stands for the pressure in the corresponding 
phase. The pressure for non-equilibrated QGP \cite{chem} can be approximated
as
\begin{equation}
P_Q(T)=\left[16 \lambda_g + 21 \lambda_q\right]\, T^4 \,\frac{\pi^2}{90}-B \, ,
\label{pq}
\end{equation}
where $\lambda_k$ is the fugacity of the parton species $k$, $B$ is the
bag-pressure, and we have considered a two flavor plasma with zero
baryochemical potential. Approximating the hadronic pressure in terms of 
a temperature dependent effective number of degrees of freedom $g_H(T)$
 ($g_H=3$, for a hadronic gas of mass-less pions, e.g.),
we can write
\begin{equation}
P_H(T)=g_H(T)\frac{\pi^2}{90} T^4 \, ,
\label{ph}
\end{equation}
which leads to
\begin{equation}
T_c^4=\frac{ 90 B/ \pi^2}{\left[16 \lambda_g + 21 \lambda_q\right]-g_H(T_c)}\, .
\label{tc}
\end{equation}

We see that if we take the bag pressure to be a constant, then decreasing
fugacities will, in general, increase the transition temperature. This is 
further accentuated by an increase in $g_H$ with temperature,
 which in fact reflects the change in the composition of the hadronic matter.
The fugacities,
measuring deviation from a chemical equilibrium,
may also have a radial dependence for a transversely expanding 
plasma~\cite{chem}.  These aspects encourage us to think of a scenario where
even the composition of the hadronic matter may become $r$ -dependent. 

Till the development of such a complete description of the collision up to
the point of freeze-out, hydrodynamics can provide a useful guideline. 
In as far as we can assume the validity of hydrodynamics, and additionally
assume an adiabatic hadronization of the plasma, we can describe the collision
in its entirety, provided that the equation of state of hadronic matter is
known.

In a recent study~\cite{crs} we addressed the question of the equation of state
in connection with the production of thermal photons and particles at 
CERN SPS in collisions involving lead nuclei, in a model calculation. 
Two approximations were used to describe the hadronic medium. 
In the first one,  only a small number of mesons 
($\pi$, $\rho$, $\omega$, and $\eta$) was  used 
(`` simplified equation of state'') while in the second one 
we used all hadrons listed in the  particle data table~\cite{PDT}.
Very different results were obtained for the transverse momentum distribution
of photons and particles, depending on the assumptions made about the
equation of state of hot hadronic matter as well as the nature of the
initially created medium.
In the present work, we show that a comparison of 
electromagnetic radiation for a number of 
final particle multiplicity densities
may help us distinguish between the equations of state. We shall also see that
the frozen-motion model, as a minimal extension~\cite{minimal}
 of  Bjorken hydrodynamics~\cite{bj}, provides us with a possibility to
 realize this distinction in an experiment done at a given energy.

A clear illustration of the effect of the equation of state on the
flow pattern at, say, LHC energies is seen in Fig.~1. We have plotted
the space-time boundaries (at $z=0$)
 of the QGP, the mixed and the 
hadronic phases, using the approach described in Ref.~\cite{crs,tr,kkmr}. 
We assume the system to be created in a state of QGP and use a boost 
invariant cylindrically symmetric transverse expansion.  We 
see that the richer equation of state leads to a shorter-lived mixed
phase and consequently the system edges past the freeze-out much more
quickly. In fact this picture holds the key to the entire discussion
that is to follow. One can, namely, immediately guess that the number of
thermal photons or thermal dileptons produced for the richer equation
of state would be smaller. One may also witness a unique sensitivity
on the extent of the strangeness equilibration~\cite{str} from such
varied flow profiles of the system. These large differences could also
be observed from pion interferometry.

In order to proceed with our discussions we consider central
collisions involving lead nuclei at a number of particle rapidity densities.
Assuming an isentropic expansion~\cite{bj} of the plasma we relate the particle
rapidity density ($dN/dy$) to the likely initial temperature ($T_i$) as
\begin{equation}
T_i^3 \, \tau_i = \frac{2\pi^4}{45\zeta (3)}\, \frac{1}{4 a_Q \pi R_T^2}
\frac{dN}{dy}\,\, ,
\end{equation}
where $a_Q=37 \pi^2/90$ for a QGP consisting of u and d quarks and gluons,
$R_T$ is the transverse dimension of the system, and we use the canonical
value of 1 fm/$c$ for the initial time $\tau_i$. We shall fix the transition
temperature at 160 MeV and the freeze-out temperature at 100 MeV.

The photon $p_T$-spectrum is obtained~\cite{us}  by convoluting
the rates for their production with the space-time history of the system as
\begin{eqnarray}
{{dN}\over {d^2p_T dy}}&=&
\int \, \tau \, d\tau \, rdr\, d\phi \,d\eta\nonumber\\
& &\left[ f_Q q_0{{dN^Q}\over {d^4xd^3q}} +
 (1-  f_Q)q_0 {{dN^H}\over {d^4xd^3q}}\right]
 \end{eqnarray}
 where the function $f_Q(r,\tau )$ denotes the fraction of the quark
 gluon plasma in the system~\cite{kkmm}. The rates in the quark and the
hadronic matter  are evaluated according to methods
 developed earlier. Thus, we have included the Compton
and the annihilation contributions from the quark matter, and the reactions
involving $\pi$, $\rho$, $\omega$ and $\eta$ mesons, along with the
$\pi \rho \rightarrow a_1 \rightarrow \pi \gamma$ reaction in 
the hadronic matter~\cite{photon}.                              

We have shown the photon multiplicities as a function of the
multiplicity of the charged particles in  Fig.~2. The slopes of the
transverse momentum distributions for the two cases were found to be quite
similar and are not shown here. We see that the ``simplified equation of
state'' for the hadronic matter leads to about twice as many photons.
This difference is likely to increase if the transition temperature
is higher than 160 MeV, which we have used, as then the difference
between the equations of state will become even more magnified.

We have also marked the likely locations of the expected values for
collisions at SPS, RHIC and  LHC energies using estimates for the
particle multiplicities from Ref.~\cite{kms}. It is quite clear that
the availability of these results will go a long way in fixing the
equation of state for the hadronic matter. From these results
we can easily see that the thermal photon multiplicity is given by,
\begin{equation}
\frac{dN_\gamma}{dy} \simeq K \left(\frac{dN_{\mathrm ch}}{dy}\right)^\alpha,
\label{scale}
\end{equation}
where $\alpha \simeq 1.2$ and $K$ depends on the equation of state.

In Fig.~3 we have repeated this analysis for the production of thermal
dileptons, using the procedures discussed in Ref.~\cite{kkmr}, for dileptons
having invariant mass $M$ between 0.5 and 1 GeV. We have included the
most dominant contribution of quark annihilation in the QGP and the
pion annihilation proceeding via the formation of a $\rho$ meson in the
hadronic matter.  We again find that
twice as many thermal dileptons are produced for the simplified equation
of state (see also Ref.~\cite{klrt}) and a scaling
similar to Eq.(\ref{scale}) with $\alpha \simeq 1.1$ is satisfied.
Apparently  the  slightly  lower value of $\alpha$ in this case
could  be  related  to  the  fact that virtual photons carrying
an invariant mass, are differently affected by the transverse flow.

At first this slower increase of the 
number of thermal photons with the charged
particle rapidity  density looks very surprising, as it is generally
believed that this dependence should be nearly
quadratic for thermal sources~\cite{czec}. It is not clear as to how
this belief got currency, as it was pointed out by Feinberg~\cite{feinberg}
 two decades ago that the number of thermal photons or dileptons
should scale as $N_\pi^{4/3}$, from very general arguments. In view of the
importance attached to this scaling behavior for identifying new 
sources of photons and dileptons it is necessary to understand it more
clearly.  Consider a system consisting of $N_{\mathrm {ch}}$ charged particles.
The number of thermal photons $N_\gamma$ will then be given by
\begin{equation}
N_\gamma~\sim~ \, e^2\, N_{\mathrm {ch}}\, \nu ~\,
\end{equation}
where $\nu$ is the number of collisions that each particle suffers.
If the system lives long enough, as when it is confined to a box,
every particle will have a chance to collide with every other particle,
and then $\nu \sim N_{\mathrm {ch}}$.
This will lead to the quadratic dependence suggested earlier~\cite{czec}.
However, the number of collisions suffered by the particles will  be 
given by $R/\lambda$, where $R$ is the size and
 $\lambda$ is the mean free path of the
particles, for a system created in heavy ion collisions~\cite{halzen}.
Realizing that the number of particles will scale as $R^3$, we immediately
get the scaling envisaged by Feinberg (see Ref.~\cite{feinberg} for arguments
when the system may be undergoing expansion, which might bring in an additional
factor of $\sqrt{\ln(N_{\mathrm {ch}})}$ ). We also realize that 
the transverse expansion of the system  drastically reduces the time
scales in the system~\cite{kkmr}. In the absence of the
transverse expansion the life-time of the system can become extremely
large; for example it will be several thousand fm/$c$, for the case discussed
in Fig.~1 (see table~I in Ref.~\cite{kkmr}).
 This would then mimic the case of 
particles contained in a box, and lead to the quadratic scaling reported by
authors of Ref.~\cite{kkmm}.

 Of course the transverse flow will considerably enhance the production of
photons having large transverse momenta, which however contribute
only marginally to the total number. This is quite a sobering thought, as
it suggests that it should be more difficult to disentangle thermal
 production from the background as  the (total) yields in both cases show a
 very similar dependence on the final state multiplicity.
 The $p_T$ distribution of the background and the thermal photons will 
thus have to play a decisive role in identifying the latter.

 If one could assume that 
the minimal extension of the Bjorken hydrodynamics invoked by a number
of authors~\cite{minimal} to evaluate the production of thermal photons and
dileptons at non-central rapidities is reliable, then one could also  
easily distinguish between different equations of state by comparing the
results at different rapidities. It will be interesting to see whether
 a more detailed hydrodynamic model
of the collision providing a fuller description at all 
rapidities~\cite{complete} can continue to support this distinction
between the two equations of state.

 It is interesting to note that the excess 
dileptons observed at SPS energies for sulphur induced collisions by 
the CERES and HELIOS experiments~\cite{ceres,helios,heros},
which are measured at different rapidities, scale as 
$\left( dN_{\mathrm {ch}}/dy \right)^\beta$, with $\beta > 1$.
The statistics, however, is still not
sufficient to distinguish the hadronic equation of state. 

In summary, we have evaluated the radiation of thermal photons and dileptons
from relativistic heavy ion collisions for a range of multiplicities. 
Assuming the formation of a quark gluon plasma in the initial state, 
and properly accounting for the transverse expansion, we find that 
thermal  production  exhibits  a  much  weaker  than  quadratic
dependence  on  the  final state multiplicity normally assumed.
It is also shown that the results are quite
sensitive to the equation of state used to describe  the hadronic matter. 
In order to distinguish and fix the equation of state we suggest to
utilize the scaling  rule of Eq.(\ref{scale}), reasonably valid for both
 real and virtual photons, as the proportionality constant is determined
by the equation of state. We are investigating whether this
constant  of proportionality is also a measure of the time-scale in the system
~Ref.\cite{halzen}.

\section*{ACKNOWLEDGEMENTS}

One of us (K.R.) acknowledges the support of the GSI Darmstadt
and of the KBN 2-P03B-09908. We thank Helmut Satz
and Bikash Sinha for useful comments.

%
%%%%%%%%%%%%% Begin figure 1 %%%%%%%%%%%%%%%%%%%%%%%%%%%%%%%%%%%%%%%%%%%%%
\begin{figure}
\psfig{figure=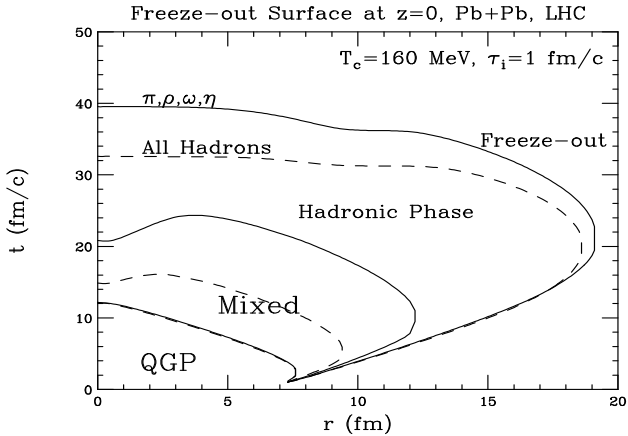,height=2.25in,width=3.25in}
\vskip 0.2cm
\caption{The space-time boundaries for the QGP phase, the mixed phase and
the hadronic phase at LHC energies. The equation of state
for the hadronic matter consists
of only $\pi$, $\rho$, $\omega$, and $\eta$ mesons (solid curves) and
of full list of hadrons in the Particle Data Book (dashed curves).}
\end{figure}
%%%%%%%%%%%%%  End figures 1 %%%%%%%%%%%%%%%%%%%%%%%%%%%%%%%%%%%%%%%%%%%%%
%%%%%%%%%%%% Begin figures 2 %%%%%%%%%%%%%%%%%%%%%%%%%%%%%%%%%%%%%%%%%%%%%
\begin{figure}
\psfig{figure=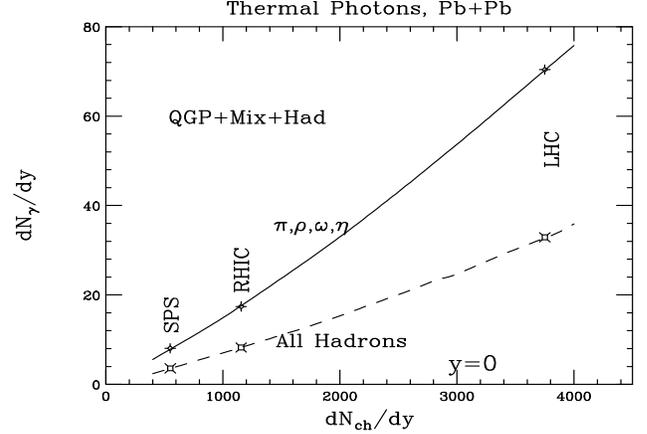,height=2.25in,width=3.25in}
\vskip 0.2cm
\caption{ Variation of rapidity density of thermal photons with the
charge particle rapidity density for the two equations of state for
the hadronic matter.}
\end{figure}
%%%%%%%%%%%%% end figures 2%%%%%%%%%%%%%%%%%%%%%%%%%%%%%%%%%%%%%%%%%%%%
%%%%%%%%%%%%% Begin figures 3 %%%%%%%%%%%%%%%%%%%%%%%%%%%%%%%%%%%%%%%%%%%%%
\begin{figure}
\psfig{figure=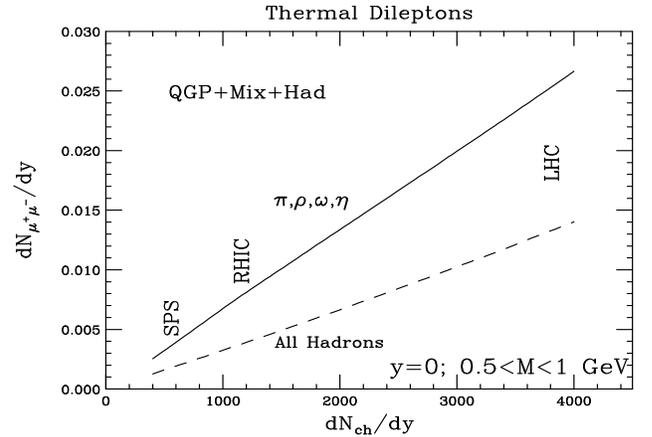,height=2.25in,width=3.25in}
\vskip 0.2cm
\caption{ Variation of rapidity density of thermal dileptons with the
charge particle rapidity density for the two equations of state for
the hadronic matter, having 0.5 $<$ $M$ 1 GeV.}
\end{figure}
%%%%%%%%%%%%% End figures 3 %%%%%%%%%%%%%%%%%%%%%%%%%%%%%%%%%%%%%%%%%%%%%

\end{document}